\documentclass[12pt,a4paper]{article}
\pdfoutput=1
\usepackage{jheppub}
\usepackage{youngtab}
\usepackage{color}
\usepackage[dvipsnames, svgnames, x11names]{xcolor}
\usepackage[utf8]{inputenc}
\usepackage{float}
\usepackage{graphicx}
\usepackage{setspace}
\usepackage{subfigure}
\usepackage{bm}
\usepackage{bbm}
\usepackage{amsmath}
\usepackage{amssymb}
\usepackage{amsbsy}
\usepackage{amsthm}
\usepackage{amsfonts}
\usepackage{braket}
\usepackage{multirow}
\usepackage{ulem}
\usepackage{slashed}
\usepackage{dsfont}
\usepackage{csquotes}
\usepackage[compat=1.1.0]{tikz-feynman}

\usepackage{enumitem}
\newlist{senum}{enumerate}{1} 
\setlist[senum]{leftmargin=4.5cm ,align=left, label=\textbf{\Roman*}} 

\def\nn{\nonumber}
\def\Tr{\text{Tr}}

\def\C{\mathcal{C}}

\def\be{\begin{equation}}       \def\ee{\end{equation}}
\def\bea{\begin{eqnarray}}      \def\eea{\end{eqnarray}}
\def\ba{\begin{array}}
	\def\ea{\end{array}}
\def\bnum{\begin{enumerate} }
	\def\enum{\end{enumerate}}

\def\nn{\nonumber}

\def\=>{\Rightarrow}
\def\>{\rightarrow}

\def\eye2{Fathbb{I}}

\def\Tr{\mathrm{Tr}}

\title{Gauge Invariants at Arbitrary $N$ and Trace Relations}

\author{Pawel Caputa$^{1,2,3}$ and}
\author{Robert de Mello Koch$^{4,5}$}

\preprint{YITP-25-137}

\affiliation[1]{The Oscar Klein Centre and Department of Physics, Stockholm University, AlbaNova, 106 91 Stockholm, Sweden}
\affiliation[2]{Yukawa Institute for Theoretical Physics, Kyoto University, Kitashirakawa Oiwakecho, Sakyo-ku, Kyoto 606-8502, Japan}
\affiliation[3]{Faculty of Physics, University of Warsaw, Pasteura 5, 02-093 Warsaw, Poland}
\affiliation[4]{School of Science, Huzhou University, Huzhou 313000, China}
\affiliation[5]{Mandelstam Institute for Theoretical Physics, School of Physics, University of the Witwatersrand, Private Bag 3, Wits 2050, South Africa}

\abstract{We investigate conformal field theories with gauge group $U(N)$ at arbitrary rank $N$, focusing on the role of trace relations in determining the structure of the Hilbert space. Working in the free trace algebra without imposing relations, we identify a class of evanescent states that vanish at finite $N$. Using the Koszul complex of \cite{Lee:2023iil}, we implement trace relations systematically via ghosts and a fermionic charge $Q_b$. This framework allows us to define and compute transition amplitudes between evanescent and physical states, which we show correspond precisely to ordinary CFT amplitudes analytically continued in $N$. Our results provide a direct algebraic realization of the proposals which realize trace relations in the bulk as over-maximal giant gravitons \cite{Gaiotto:2021xce,Lee:2023iil,Lee:2024hef} and establish analytic continuation in $N$ as a powerful tool for understanding finite-$N$ effects.}
\begin{document}
\sloppy
\maketitle

\section{Introduction}
The study of conformal field theories (CFTs) with gauge group $U(N)$ is central to our understanding of holography and the AdS/CFT correspondence. At large $N$, correlators simplify and the Hilbert space organizes naturally in terms of single- and multi-trace operators. Different trace structures are orthogonal, and the basis of operators generated by single traces admits a direct identification with the Fock space of supergravity. At finite $N$, however, the situation becomes far more subtle: trace structures mix and the trace basis no longer diagonalizes the two-point function. Remarkably, an alternative basis exists, given by Schur polynomials~\cite{Corley:2001zk}, which continues to diagonalize the free two-point function even at finite $N$. This construction generalizes to gauge groups $O(N)$ and $Sp(N)$~\cite{Caputa:2013hr,Caputa:2013vla}, and at finite $N$ it naturally encodes non-perturbative brane states such as giant gravitons ~\cite{Balasubramanian:2001nh,Corley:2001zk,Aharony:2002nd}, as well as the $\tfrac{1}{2}$-BPS Lin-Lunin-Maldacena (LLM) geometries ~\cite{Lin:2004nb}. Together with natural multi-matrix extensions~\cite{Kimura:2007wy,Brown:2007xh,Bhattacharyya:2008rb,Bhattacharyya:2008xy,Brown:2008ij}, this basis provides a powerful framework for computing CFT correlators involving giant gravitons and LLM geometries~\cite{Berenstein:2004kk,Balasubramanian:2005mg,Berenstein:2017rrx,deMelloKoch:2008ugi,Bissi:2011dc,Caputa:2012yj,Lin:2012ey,Garner:2014kna}. A further complication at finite $N$ is that multi-trace operators are not independent but instead satisfy non-trivial trace relations \cite{procesi1976invariant}. These relations underlie important phenomena such as the stringy exclusion principle and the truncation of Hilbert spaces \cite{deMelloKoch:2025ngs,deMelloKoch:2025rkw}. Understanding these finite-$N$ effects is therefore essential for probing physics beyond the planar approximation.

In this work we explore a complementary perspective: instead of fixing $N$ from the outset, we consider $U(N)$ CFTs at arbitrary $N$. Mathematically, this corresponds to working with the free trace algebra in which no relations are imposed. In this enlarged Hilbert space, one finds states that have no counterpart at finite $N$; we refer to these as evanescent states, since they vanish once the trace relations are enforced. These evanescent states are closely related to states of the same name that appear (for example) in studies of the Wilson-Fisher CFT when $d$ is continued to non-integer values~\cite{Hogervorst:2015akt}. Our construction provides a universal object that interpolates between all finite-$N$ theories and offers a natural setting for analytic continuation in $N$. Classical $\frac{1}{16}$-BPS states in super Yang–Mills theory at non-integer gauge group rank—which is closely related to the present analysis—were recently studied in \cite{Budzik:2023vtr}.

Our work is closely connected to recent developments that probe the physics of trace relations directly in the bulk gravity dual~\cite{Gaiotto:2021xce,Lee:2023iil,Lee:2024hef}.  This line of research has led to several concrete results. First, by analyzing the role of trace relations, the giant graviton expansion was discovered~\cite{Gaiotto:2021xce}. See~\cite{Imamura:2021ytr,Murthy:2022ien,Lee:2023iil,Lee:2024hef,Liu:2022olj,Eniceicu:2023uvd,Chen:2024cvf}, for closely related work. We will show that our description in terms of evanescent states reproduces an identity that is mathematically identical to the giant graviton expansion. Second, trace relations in the bulk have been related to ghost states, together with a fermionic charge $Q_b$ whose cohomology selects the physical Hilbert space~\cite{Lee:2023iil}. Third, \cite{Lee:2024hef} developed a thimble formalism in which the ghost states dual to trace relations are realized as specific thimbles entering the path integral of a maximal giant. Within our framework, we propose that the ghost brane states\footnote{The phrase ``ghost branes'' has appeared previously in the literature to describe D-branes of negative action~\cite{Okuda:2006fb}. In this paper, our use of the term is unrelated to that context. By ``ghost brane states'' we specifically mean states associated with giant graviton D3-branes, defined using a rotated contour prescription. These states serve as ``ghosts'' for null states that appear in the full Hilbert space of the theory, and their introduction provides a convenient device for organizing and resolving redundancies in the spectrum~\cite{Gaiotto:2021xce,Lee:2023iil,Lee:2024hef}.} correspond to evanescent states, and further, that transition amplitudes involving ghost brane states and ordinary branes or gravitons are precisely given by ordinary CFT amplitudes analytically continued in $N$. The identification of ghost brane states with evanescent states is natural, since the latter vanish once trace relations are imposed. We illustrate these ideas with explicit examples, focusing on the half-BPS sector where connections to giant gravitons and the stringy exclusion principle are most transparent.

Taken together, our results establish a concrete bridge between analytically continued $U(N)$ CFTs and the framework developed in~\cite{Gaiotto:2021xce,Lee:2023iil,Lee:2024hef}. More broadly, they offer a new perspective on the role of finite-$N$ constraints in holography and underscore the utility of analytic continuation in $N$ as a tool for probing the physics of gauge-invariant operator algebras.

The paper is organized as follows. In Section~\ref{sec:schurholog} we review the Schur polynomial basis and its role in realizing finite $N$ physics of holographic descriptions of the $U(N)$ CFT. Section~\ref{sec:EvanescentStates} introduces the analytic continuation in $N$ of the CFT and develops the giant graviton-like expansion. In Section~\ref{sec:TransitionAmplitudes} we present our proposal for transition amplitudes between ghost brane states and states corresponding to physical giant gravitons. Section~\ref{sec:Conclusions} contains a discussion of our results and suggests how this work might be extended.

\section{Holography of the $\frac12$-BPS Sector at finite $N$}\label{sec:schurholog}
The $\tfrac{1}{2}$-BPS sector is remarkably rich: it contains all the essential ingredients needed to probe finite-$N$ physics. In this sector we encounter giant graviton branes, whose angular momentum is bounded above by $N$, as well as the LLM geometries, which arise from condensates of giant gravitons. These features make the $\tfrac{1}{2}$-BPS sector an ideal laboratory for exploring finite-$N$ effects. Our focus will be on the CFT description, where finite-$N$ physics manifests in correlation functions evaluated to all orders in $1/N$. In later sections we will extend this description to non-integer $N$. To prepare for that, the present section has two main goals:
\begin{itemize}
\item[1.] To show that the $N$-dependence of correlation functions generically enters through arguments of $\Gamma(\cdot)$ functions. This is significant because the analytic continuation of $\Gamma(\cdot)$ is standard and well understood, making it a natural tool for continuing correlators to arbitrary $N$.
\item[2.] To establish the precise match between CFT correlators and predictions of the dual gravity theory. This agreement is non-trivial, as it holds at finite $N$, far from the planar regime where weakly coupled string theory applies. A particularly clear case is provided by giant gravitons, which appear as solitonic objects in the bulk. We will show that transitions between distinct soliton states are exactly reproduced by heavy-heavy-light correlators in the CFT.
\end{itemize}
We begin with a brief review of the background needed to compute correlators in the CFT using Schur polynomials. Specifically, we recall how two- and three-point functions of $\tfrac{1}{2}$-BPS Schur operators in $\mathcal{N}=4$ SYM map to processes involving giant gravitons in $AdS_5 \times S^5$. This already illustrates that $N$-dependence consistently enters through $\Gamma(\cdot)$. We then show that these CFT correlators precisely reproduce the transition amplitudes between giant gravitons in the dual gravity description, first using the DBI brane world-volume action and then via the supergravity construction of~\cite{Lin:2004nb}. This establishes a detailed and exact agreement between CFT and gravity at finite $N$. While much of this section reviews earlier work~\cite{Bissi:2011dc,Caputa:2012yj,Lin:2012ey,Bissi:2013qmo}, there are some new results. We introduce a novel class of generalized Schur operators. These operators are dual to giant gravitons localized at position $\lambda$ on the LLM plane, with $\lambda$ serving as the parameter of the generating function. These provide generating functions for trace relations and allow us to evaluate the complete set of correlators among them.
\subsection{Correlators with Schurs}
Consider the  Schur polynomial basis \cite{Corley:2001zk} for N=4 SYM with $U(N)$ gauge group\footnote{For discussion on how to extend this technology to $SU(N)$ gauge group see \cite{deMelloKoch:2004crq}.}
\be
\chi_R(Z)=\frac{1}{n!}\sum_{\sigma\in S_n}\chi_R(\sigma)\Tr(\sigma Z^{\otimes n})\,,
\ee
where $Z=\phi^1+i\phi^2$ is a complex scalar built from two (out of six) scalars of the model. R stands for a Young diagram with n boxes, that is, an irreducible
representation of the symmetric group $S_n$, $\sigma$ labels a particular permutation and $\chi_R(\sigma)$ is its character. These operators are orthogonal and form a basis in the 1/2 BPS sector at finite $N$. Their two-point correlators (suppressing the usual CFT spacetime dependence) are
\be
\langle \chi_R(Z)\chi_S(\bar{Z})\rangle=\delta_{R,S}f_R(N)\,,\label{eq:2pSchur}
\ee
where $f_R(N)$ is the product of the weights of $R$, which for the box in the $i^{th}$ row and the $j^{th}$ column of the Young diagram are given by $N-i+j$. The quantity $f_R(N)$, is thus a product of factors, each of which is a product of consecutive integers. Any product of consecutive integers can be expressed as a ratio of $\Gamma(\cdot)$ functions. Clearly then, $f_R(N)$ can always be expressed in terms of ratios of $\Gamma(\cdot)$ functions. Since the analytic continuation of the $\Gamma(\cdot)$ function is well understood, the correlation functions of Schur polynomial are easily continued to arbitrary complex values of $N$.

\begin{figure}[h!]
\hspace{0.3cm}
\includegraphics[width=.46\textwidth]{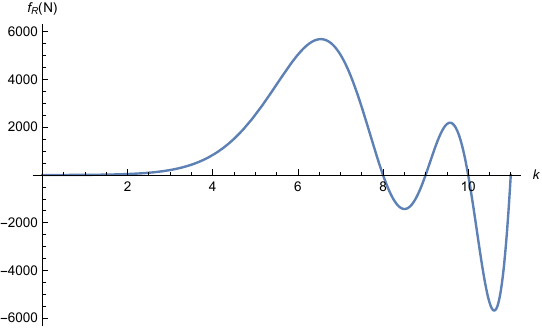}
\hspace{0.3cm}
\includegraphics[width=.46\textwidth]{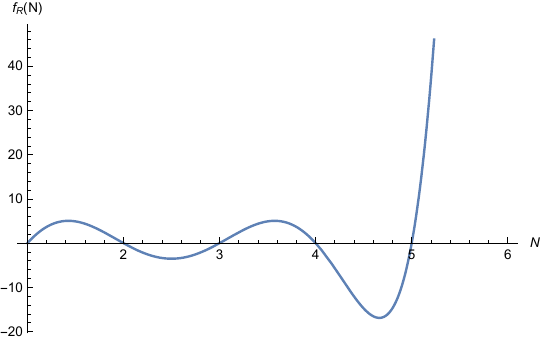}
\caption{Plots of the two-point correlator \eqref{eq:fh1jN} for sub-determinants labelled by a single column with k boxes $(1^k)$, as function of $k$ (left for $N=7$) and $N$ (right for $k=6$).}
\label{fig:2pt}
\end{figure}

In the following, we focus on the class of so-called hook diagrams $h^j_k$ which have $k$ boxes, with the first row of length $j$ and $k-j$ rows of length one. For these we have
\be
f_{h^{j}_k}(N)=\prod^{j}_{i=1}(N+i-1)\prod^{k-j}_{i=1}(N-i)=\frac{\Gamma(N+j)}{\Gamma(N-k+j)}\,.\label{eq:fhkjN}
\ee
A simple family of hooks is given by a single column with k boxes that we denote by
\be
h^{1}_k\equiv(1^k)\,,
\ee
and for this class we have
\be
f_{(1^k)}(N)=\frac{N!}{(N-k)!}=\frac{\Gamma(N+1)}{\Gamma(N-k+1)}\,.\label{eq:fh1jN}
\ee
It is instructive to plot \eqref{eq:fh1jN} as functions of $N$ or $k$, as shown in Fig.~\ref{fig:2pt}. Clearly, they depend non-trivially on these parameters but vanish when $k>N$ with integer $N$.

Operators labelled by $(1^k)$ are often called sub-determinants. The name comes from the fact that for $k=N$ they reduce to determinant of $Z$ 
\be
\chi_{(1^N)}(Z)=\det(Z)\,.
\ee
Schur polynomials also satisfy the product rule
\be
\chi_R(Z)\chi_S(Z)=\sum_{T}g(R,S;T)\chi_T(Z)\,,\label{LRRule}
\ee
where $R$, $S$ and $T$ have $k_1$, $k_2$, and $k_1+k_2$ boxes respectively, and $g(R,S;T)$ stands for the Littlewood-Richardson coefficient (multiplicity). For our computations, we will need the fact that when composing two hook diagrams of size $h^{j_1}_{k_1}$ and $h^{j_2}_{k_2}$ we can obtain hooks $h^{j_1+j_2}_{k_{1}+k_2}$ and $h^{j_1+j_2-1}_{k_{1}+k_2}$, both with multiplicities 1, as well as additional terms whose details form is not important as they do not contribute, see \cite{Caputa:2012dg}). The product rule allows us to reduce the computation of three-point (3pt) correlators to the computation of a two point correlator. For sub-determinants, we have (again suppressing the universal spacetime dependence of 3pt functions in CFTs)
\be
\langle \chi_{(1^{k_1})}(Z)\chi_{(1^{k_2})}(Z)\chi_{(1^{k_1+k_2})}(\bar{Z})\rangle=f_{(1^{k_1+k_2})}(N)=\frac{\Gamma(N+1)}{\Gamma(N-k_1-k_2+1)}\,.
\ee

Another set of operators that serve as useful probes are the single-trace chiral-primary operators
\be
\Tr(Z^J)=\sum_R\chi_R(\sigma_J)\chi_R(Z)=\sum^J_{k=1}(-1)^{J-k}\chi_{h^k_{J}}(Z)\,,\label{eq:2pCPO}
\ee
where in the second step we moved to the Schur polynomial basis (where $\sigma_J$ stands for any $J$-cycle) and in the third step we used the fact that the characters of $J$-cycles are non-zero only for hooks where they read $\chi_{h^k_{J}}(\sigma_J)=(-1)^{J-k}$ (see detailed discussion in \cite{Corley:2002mj,Caputa:2012dg}). This way, the two-point correlator (again suppressing the spacetime dependence) becomes
\be
\langle \Tr(Z^J)\Tr(\bar{Z}^J)\rangle=\sum^J_{j=1}f_{h^{j}_J}(N)=\frac{1}{J+1}\left(\frac{\Gamma(N+J+1)}{\Gamma(N)}-\frac{\Gamma(N+1)}{\Gamma(N-J)}\right)\,.
\ee
The rules above are sufficient to compute correlators between the Schurs and the chiral primaries. We will only need this result for sub-determinants where the answer reads
\be
\langle \chi_{(1^k)}(Z)\Tr(\bar{Z}^J)\rangle=\delta_{k,J}(-1)^{k-1}f_{(1^k)}(N)\,.\label{CPOSchur2pt}
\ee
This way, we can derive the 3pt correlator
\bea
\langle \Tr(Z^J)\chi_{(1^{k-J})}(Z)\chi_{(1^{k})}(\bar{Z})\rangle&=&\sum^J_{j=1}(-1)^{J-j}g(h^j_J,(1^{k-J});(1^k))f_{(1^k)}(N)\nn\\
&=&(-1)^{J-1}f_{(1^k)}(N)\,.
\eea
Finally, we put all these ingredients together to compute normalized three point correlators of two Schur polynomials and one chiral-primary \cite{Bissi:2011dc,Caputa:2012yj,Lin:2012ey} 
\be
C^{(3)}\equiv\frac{\langle \Tr(Z^J)\chi_{(1^{k-J})}(Z)\chi_{(1^k)}(\bar{Z})\rangle}{\sqrt{\langle \Tr(Z^J)\Tr(\bar{Z}^J)\rangle\langle \chi_{(1^{k-J})}(Z)\chi_{(1^{k-J})}(\bar{Z})\rangle\langle \chi_{(1^k)}(Z)\chi_{(1^{k})}(\bar{Z})\rangle}}\,,
\ee
where each of the operators was divided by its norm. These correlators are usually refereed to as extremal, due to the fact that the sum of conformal dimensions of the operators in matrix $Z$ equals to the sum dimensions of operators in $\bar{Z}$ i.e., $J+(k-J)=k$. The answer reads \cite{Bissi:2011dc}
\bea
C^{(3)}&=&(-1)^{J-1}\sqrt{\frac{\Gamma(N-k+J-1)}{\frac{\Gamma(N-k+1)}{J+1}\left(\frac{\Gamma(N+J+1)}{\Gamma(N)}-\frac{\Gamma(N+1)}{\Gamma(N-J)}\right)}}\,.\label{GenForm}
\eea

These three point correlators can be interpreted as a process where the (light) chiral-primary operator probes (heavy) multi-trace Schur polynomial operators. This can be made very precise by taking the Heavy-Heavy-Light (HHL) limit of the three point correlator where 
\be
N\to\infty,\qquad k\to\infty,\qquad \frac{k}{N}<1-\text{finite}\,. \label{HHLLimit}
\ee
In this limit we can approximate \eqref{eq:2pCPO} by $JN^J$ and the HHL 3pt function becomes 
\bea
C^{(3)}_{HHL}\simeq(-1)^{J-1}\sqrt{\frac{\prod^{k}_{j=k-J+1}N(1-\frac{j-1}{N})}{JN^J}}
=\frac{(-1)^{J-1}}{\sqrt{J}}\left(1-\frac{k}{N}\right)^{J/2}\,.\label{3ptAHHL}
\eea
We will return to this result in \ref{subsec:HHL} and reproduce it from the dual gravity.
\subsection{Generalized sub-determinants}
We now introduce a generalization of the sub-determinant operators that directly make contact with the trace relations. Towards this end, consider a continuous family of sub-determinants $\chi_{(1^k)}(\lambda\mathbf{1}-Z)$ where $\lambda\in\mathbb{C}$. As we will discuss in \ref{subsec:HHL}, sub-determinant operators are dual to giant gravitons. The generalization with parameter $\lambda$ shifts the eigenvalues of $Z$ and then evaluates the subdeterminant. To derive a useful representation of these operators, let us first recall the explicit form of sub-determinant operators for several low values of $k$
\bea 
\chi_{(1)}(Z)&=&\Tr(Z)\,,\nn\\
\chi_{(1^2)}(Z)&=&\frac{1}{2}\left(\Tr(Z)^2-\Tr(Z^2)\right)\,,\nn\\ \chi_{(1^3)}(Z)&=&\frac{1}{6}\left(2\Tr(Z^3)-3\Tr(Z)\Tr(Z^2)+\Tr(Z)^3\right)\,,\nn\\
\chi_{(1^4)}(Z)&=&\frac{1}{24}\left(\Tr(Z)^4-6\Tr(Z^2)\Tr(Z)^2+3\Tr(Z^2)^2+8\Tr(Z)\Tr(Z^3)-6\Tr(Z^4)\right)\,.\nn\\
\eea
Replacing matrix $Z\to \lambda\mathbf{1}-Z$, and using $\Tr(1)=N$, leads to an explicit expansion (see similar expansions for $SU(N)$ in \cite{deMelloKoch:2004crq})
\bea 
\chi_{(1)}(\lambda\mathbf{1}-Z)&=&\lambda N-\Tr(Z)=\lambda N-\chi_{(1)}(Z)\,,\nn\\
\chi_{(1^2)}(\lambda\mathbf{1}-Z)&=&\frac{N(N-1)}{2}\lambda^2-(N-1)\lambda \chi_{(1)}(Z)+\chi_{(1^2)}(Z)\,,\nn\\ \chi_{(1^3)}(\lambda\mathbf{1}-Z)&=&\frac{N(N-1)(N-2)}{6}\lambda^3-\frac{(N-1)(N-2)}{2}\lambda^2\chi_{(1)}(Z)\nn\\
&+&(N-2)\lambda\chi_{(1^2)}(Z)-\chi_{(1^3)}(Z)\,,\nn\\
\chi_{(1^4)}(\lambda\mathbf{1}-Z)&=&\frac{N(N-1)(N-2)(N-3)}{24}\lambda^4-\frac{(N-1)(N-2)(N-3)}{6}\lambda^3\chi_{(1)}(Z)\nn\\
&+&\frac{(N-2)(N-3)}{2}\lambda^2\chi_{(1^2)}(Z)-(N-3)\lambda \chi_{(1^3)}(Z)+\chi_{(1^4)}(Z)\,.
\eea
This allows us write the generalized operators as
\be
\chi_{(1^k)}(\lambda\mathbf{1}-Z)=\sum^k_{j=0}(-1)^j\binom{N-j}{k-j}\lambda^{k-j}\chi_{(1^j)}(Z)\,.
\ee
Again, it is natural to rewrite the binomial coefficient in terms of Gamma functions to obtain
\be
\binom{N-j}{k-j}=\frac{\Gamma(N-j+1)}{\Gamma(N-k+1)\Gamma(k-j+1)}\,.
\ee
Note that, irrespective of $N$, the coefficient $\Gamma(k-j+1)$ vanishes for $j>k$, bearing in mind that $k$ is an integer. Clearly, the special operator in this family is the determinant with $k=N$, for which the binomial coefficient becomes 1 and we find
\be
\det(\lambda\mathbf{1}-Z)=\sum^k_{j=0}(-1)^j\lambda^{k-j}\chi_{(1^j)}(Z)\,.
\ee
Interestingly, for determinants, there is an alternative derivation of this result as follows. For $N\times N$ matrix $Z$, the determinant can be written as
\be
\det(\lambda\mathbf{1}-Z)=\det(\lambda\mathbf{1})\det(\mathbf{1}-Z/\lambda)=\lambda^Ne^{\Tr\log(\mathbf{1}-Z/\lambda)}\,.
\ee
Then, using the series expansion of the logarithm
\be
\log(1-x)=-\sum^\infty_{j=1}\frac{x^j}{j}\,,
\ee
we can write the above operator as
\be
\det(\lambda\mathbf{1}-Z)=\lambda^Ne^{-\sum^\infty_{j=1}\frac{\lambda^{-j}}{j}\Tr(Z^j)}\equiv \sum^\infty_{j=0}(-1)^j\lambda^{N-j}\chi_{(1^j)}(Z)\,.\label{DetGen}
\ee
Observe that the sum in the second derivation extends all the way to infinity. However, for integer $N$, all the operators $\chi_{(1^j)}(Z)$ with $j>N$ vanish, and in fact define the trace relations \cite{procesi1976invariant}. Hence, to encompass the entire family of the generalized sub-determinants, we formally write
\be
\chi_{(1^k)}(\lambda\mathbf{1}-Z)\equiv\sum^\infty_{j=0}(-1)^j\frac{\Gamma (N-j+1)}{\Gamma (k-j+1) \Gamma (N-k+1)}\lambda^{k-j}\chi_{(1^j)}(Z)\,.
\ee

In order to see the role played by this formal expansion consider the two-point correlator of determinants 
\be
\langle \det(\lambda_1\mathbf{1}-Z) \det(\lambda_2\mathbf{1}-\bar{Z})\rangle=\lambda^N_1\lambda^N_2\left(\sum^N_{k=0}+\sum^\infty_{k=N+1}\right)\frac{1}{\lambda^k_1\lambda^k_2}\frac{\Gamma(N+1)}{\Gamma(N-k+1)}\,,
\ee
where the sum in \eqref{DetGen} is separated into terms with $k$ up to $N$ and those with $k$ exceeding $N$. The first sum gives incomplete $\Gamma$ function so we can write it as
\be
\langle \det(\lambda_1\mathbf{1}-Z) \det(\lambda_2\mathbf{1}-\bar{Z})\rangle=e^{\lambda_1\lambda_2}\Gamma(N+1,\lambda_{1}\lambda_2)+\sum^\infty_{k=N+1}\frac{\lambda^N_1\lambda^N_2}{\lambda^k_1\lambda^k_2}\frac{\Gamma(N+1)}{\Gamma(N-k+1)}\,.
\ee
Using Euler's reflection formula
\be
\Gamma(-z)=\frac{1}{\Gamma(z+1)}\frac{\pi}{\sin\left(\pi(z+1)\right)}\,,
\ee
we can rewrite the Gamma function as
\be
\Gamma(N-k+1)=\Gamma(-(k-N-1))=\frac{1}{\Gamma(k-N)}\frac{\pi}{\sin\left(\pi(k-N)\right)}\,,
\ee
and the correlator becomes
\bea
\langle \det(\lambda_1\mathbf{1}-Z) \det(\lambda_2\mathbf{1}-\bar{Z})\rangle&=&e^{\lambda_1\lambda_2}\Gamma(N+1,\lambda_{1}\lambda_2)\nn\\
&+&\frac{\Gamma(N+1)}{\pi}\sum^\infty_{k=N+1}\frac{\Gamma(k-N)\sin\left(\pi(k-N)\right)}{(\lambda_1\lambda_2)^{k-N}}\,.\nn\\
\eea
Notice that for integer $N$ the second term on the RHS vanishes. We nevertheless keep it here as it will play a role after we continue to non-integer $N$.
\subsection{Correlators with generalized sub-determinants}
Using the rules introduced in the previous section, we can compute the correlators between generalized sub-determinant operators as well as trace relations. Let us first start with the sub-determinant of dimension $k$ and a single-trace chiral-primary
\bea
\langle\chi_{(1^k)}(\lambda\mathbf{1}-Z)\Tr(\bar{Z}^J)\rangle&=&\sum^\infty_{j=0}(-1)^j\binom{N-j}{k-j}\lambda^{k-j}\langle\chi_{(1^j)}(Z)\Tr(\bar{Z}^J) \rangle\nn\\
&=&-\lambda^{k-J}\frac{\Gamma(N+1)}{\Gamma(N-k+1)\Gamma(k-J+1)}\,,
\eea
where in the second step we have used \eqref{CPOSchur2pt}. Since $k$ is an integer, this correlator vanishes for $J>k$. We can first consider the special case of $k=N$ to find
\be
\langle \det(\lambda\mathbf{1}-Z) \Tr(\bar{Z}^J)\rangle=-\lambda^{N-J}\frac{\Gamma(N+1)}{\Gamma(N-J+1)}\,.\label{CPtracerel}
\ee
We discuss this correlation function further in Section \ref{transitions}.

Next, lets compute the two-point correlator of a generalized sub-determinant operator with the standard Schur polynomial
\bea
\langle \chi_{(1^k)}(\lambda\mathbf{1}-Z)\chi_{R}(\bar{Z})\rangle &=&\sum^\infty_{j=0}(-1)^j\binom{N-j}{k-j}\lambda^{k-j}\delta_{R,(1^j)}f_{R}(N)\,.
\eea
This correlator of Schur operators is non-trivial only for single column of size $n$ $R=(1^n)$ where it reads
\bea
\langle \chi_{(1^k)}(\lambda\mathbf{1}-Z)\chi_{(1^n)}(\bar{Z})\rangle&=&\frac{(-1)^n\lambda^{k-n}\,\Gamma(N+1)}{\Gamma(k-n+1)\Gamma(N-k+1)}\,.\label{DetR}
\eea
This is consistent with the fact that generalised sub-determinants form an algebraic basis in terms of which we can write an arbitrary Schur polynomial (see Appendix A of \cite{deMelloKoch:2007rqf} and \cite{ledermann1987introduction}). We return to this formula in Section \ref{transitions}.
\subsection{HHL three-point functions in AdS/CFT}\label{subsec:HHL}
We will now briefly review how to reproduce the semi-classical HHL correlators \eqref{3ptAHHL} from semi-classical gravity computations. This line of research was started in \cite{Janik:2010gc,Zarembo:2010rr,Costa:2010rz} and computations of HHL correlators with giant gravitons were first done in \cite{Bissi:2011dc}. Symmetry realization and giant graviton state dependence is analyzed in \cite{Bajnok:2014sza,Yang:2021kot}, while subtleties specific to extremal states are examined in \cite{Holguin:2022zii}.
\subsubsection{HHL correlators from DBI action}
The gravity dual of the 1/2 BPS sub-determinant operator $\chi_{(1^k)}(Z)$ is described by the probe D3 brane wrapping $S^3\subset S^5$ \cite{McGreevy:2000cw,Grisaru:2000zn,Hashimoto:2000zp,Corley:2001zk,Balasubramanian:2001nh}. The Lorentzian action describing a single D3 brane is the standard (bosonic) DBI (see \cite{Das:2000st} for useful background)
\bea
S^{DBI}_{(3)}&=&-T_3\int d^{4}\sigma\sqrt{-g}+T_3\int P[C^{(4)}]\,,
\eea
where $g$ is the pull-back of the spacetime metric to the world-volume of the brane and $P[C^{(4)}]$ is the pull-back of the 4-form flux. The relation between the brane tension $T_3$ and the units of the flux is fixed by supergravity equations and is given by
\be
T_3 A_3=N\,,
\ee
where $A_3=2\pi^2$ is the area of a unit three-sphere. Now, consider the metric on $AdS_5\times S^5$
\be
ds^2=-\cosh^2\rho\,dt^2+d\rho^2+\sinh^2\rho\, d\tilde{\Omega}^2_3+d\theta^2+\sin^2\theta \,d\phi^2+\cos^2\theta\, d\Omega^2_3\,,
\ee
in which the embedding of the D3 is described by following ansatz
\be
\rho=0\,,\qquad \sigma^0=t\,,\qquad \theta=\text{const.}\,,\qquad \phi=\phi(t)\,,\qquad \sigma^i=\chi_i\,,
\ee
where $\chi_i$ are coordinates on $S^3\in S^5$.
With this ansatz, the action
\be
S^{DBI}_{(3)}=-N\int dt\left(\sqrt{1-\sin^2\theta\, \dot{\phi}(t)^2}\cos^3\theta-\dot{\phi}(t)\cos^4\theta\right)\,,
\ee
has a conserved angular momentum 
\be
k=\frac{\partial L}{\partial \dot{\phi}(t)}=\frac{N\dot{\phi}(t)\sin^2\theta\cos^3\theta}{\sqrt{1-\sin^2\theta\, \dot{\phi}(t)^2}}+N\cos^4\theta\,,
\ee
so that we can find the relevant saddle by minimizing the Hamiltonian (Routhian)
\be
\mathcal{H}=\dot{\phi}k-L=\frac{N}{\sin\theta}\sqrt{(l-\cos^4\theta)^2+\sin^2\theta\cos^6\theta}\,,
\ee
where $l=k/N$. 
\begin{figure}[b!]
\hspace{0.3cm}
\includegraphics[width=.45\textwidth]{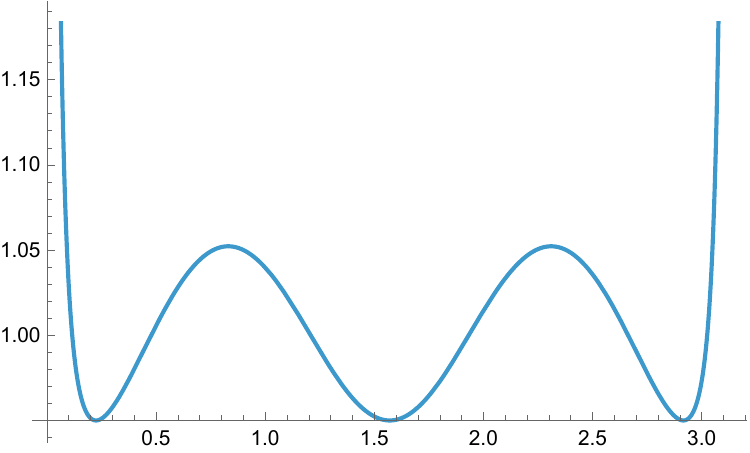}
\hspace{0.3cm}
\includegraphics[width=.45\textwidth]{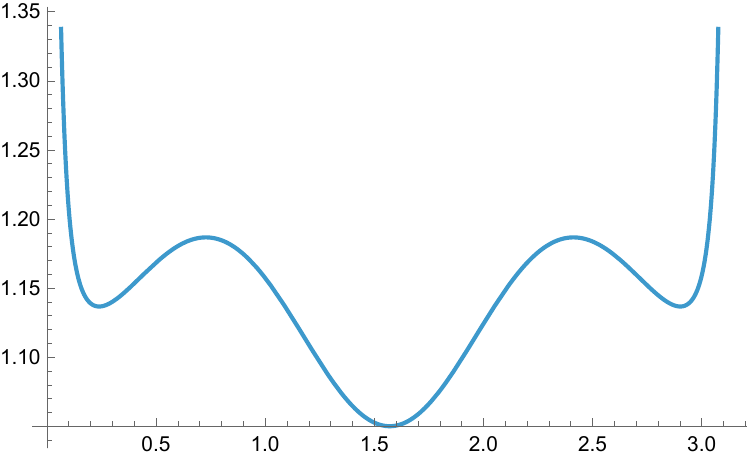}
\caption{Plots of $\mathcal{H}/N$ for $l<1$ (left) and $l>1$ (right). Here we chose $l=19/20$ and $l=21/20$. }
\label{fig:Hl}
\end{figure}
For $l<1$, we have three minima (left figure on Fig. \ref{fig:Hl}) at
\be
\theta=\{\arccos(\sqrt{l}),\frac{\pi}{2},\pi-\arccos(\sqrt{l})\}\,,
\ee
and at each of them
\be
\cos^2\theta=\frac{k}{N}\,,\qquad \mathcal{H}=Nl=k\,.\label{costhetakN}
\ee
It turns out that the stable solution with $\theta=\pi/2$, called the giant-graviton, is associated with sub-determinant $\chi_{(1^k)}(Z)$ and its angular momentum $k$ is translated to the number of boxes of the Young diagram labelling the CFT operator.

It is also interesting to point that the maxima of $\mathcal{H}$ have the value
\be
\cos^2\theta=\frac{3-\sqrt{9-8\frac{k}{N}}}{4}\,.
\ee
Moreover, for $l>1$, two of the three minima are lifted becoming unstable (right figure on Fig. \ref{fig:Hl}).
Their dependence on $\theta$ is given by 
\be
\cos^2\theta=\frac{3+\sqrt{9-8\frac{k}{N}}}{4}\,.
\ee

Finally, following \cite{Janik:2010gc,Zarembo:2010rr,Costa:2010rz}, the semi-classical HHL 3pt correlators can be reproduced by varying the Euclidean, on-shell DBI action evaluated on the giant graviton, with appropriate mode corresponding to the CFT operator. The giant graviton solutions described here carry two non-vanishing charges which break dilatation symmetry and $R$-charge rotations. As explained in \cite{Bajnok:2014sza,Yang:2021kot} we must perform an orbit average over classical solutions to restore these symmetries. In general, the fluctuation is given by \cite{Bissi:2011dc,Lin:2012ey}
\be
\delta S^{DBI}_{(3)}=N\cos^2\theta
\left[\frac{2}{\Delta+1}Y_\Delta(\partial^2_t-\Delta^2)s^\Delta+4(\Delta \cos^2\theta-\sin\theta\cos\theta\partial_\theta)Y_\Delta s^\Delta\right]_{t=0}\,,
\ee
where the bulk to boundary propagator is
\be
s^\Delta=\frac{a_\Delta z^\Delta}{((x-x_B)^2+z^2)^\Delta}
\,,\qquad a_\Delta=\frac{\Delta+1}{N\,2^{2-\frac{\Delta}{2}}\sqrt{\Delta}}\,,
\ee
and $Y_\Delta$ is the spherical harmonic representing the supergravity mode dual to $\Tr(Z^J)$ 
\be
Y_\Delta=\frac{\sin^\Delta\theta}{2^{\Delta/2}}e^{i\Delta \phi}\,,\qquad \Delta=J\,.\label{SHTrZJ}
\ee
However, there is an additional well-known subtlety in the computation of the extremal correlators~\cite{DHoker:1999jke}, where the sum of dimensions of operators in matrix $Z$ equals to the dimension of operators in $\bar{Z}$ (such as e.g. \eqref{3ptAHHL}). To reproduce the extremal HHL 3pt functions one can use a regularisation trick~\cite{DHoker:1999jke,Lin:2012ey} and compute the correlator of two heavy sub-determinants with a light operator \footnote{The spherical harmonic corresponding to this operator is
\be
Y_{\Delta,\Delta-2l}=\frac{\Gamma(J+l+1)\sqrt{(l+1)(J+l+1)}}{2^{l+J/2}\Gamma(l+2)\Gamma(J+1)\sqrt{J+2l+1}}\sin^J\theta\,e^{iJ\phi}\,_2F_1(-l,J+l+2,J+1;\sin^2\theta)\,,
\ee
and in the limit of $l\to 0$ reproduces \eqref{SHTrZJ}.}
\be
\Tr(Z^{J+l}\bar{Z}^l)\,,
\ee
and then take the limit of $l\to 0$. The final answer for the fluctuation of the DBI action in this case reads \cite{Lin:2012ey}
\be
\frac{\delta S^{DBI}_{(3)}}{(2\mathcal{R})^\Delta}= \frac{\Gamma(J+l+1)\Gamma(J+l)\sqrt{(l+1)(J+l+1)}}{\Gamma(J+2l)\Gamma(J+1)\sqrt{J+2l+1}\sqrt{j+2l}}\sin^J\theta\,_2F_1(-l,J+l,J+1;\sin^2\theta)\,,
\ee
where $\mathcal{R}$ is a factor that depends on the spatial separation $x_B$ (see \cite{Bissi:2011dc}). Finally, in the limit of $l\to0$ we simply get
\be
\frac{\delta S^{DBI}_{(3)}}{(2\mathcal{R})^\Delta}=\frac{\sin^J\theta}{\sqrt{J}}=\frac{1}{\sqrt{J}}\left(1-\frac{k}{N}\right)^{J/2}\,,
\ee
where in the last step we used \eqref{costhetakN} reproducing the extremal correlator \eqref{3ptAHHL}, up to the $(-1)^{J-1}$ factor that can be added/removed by an appropriate choice of convention~\cite{Bissi:2011dc}. As emphasized in \cite{Holguin:2022zii}, the apparent divergence is removed without regularization by first carrying out the orbit average which restores $R$-symmetry.

Observe that the fluctuation of $S^{DBI}_{(3)}$ is relatively universal in terms of its dependence on $\theta$. Indeed we only needed the relation between $\theta$ and $k/N$ at the very end. This way, it is natural to expect that one may be able to reproduce correlators with trace relations by similar analysis continued to appropriate regions of the parameter space. We leave this for the future.
\subsubsection{HHL correlators from LLM geometries}
There exists an equivalent method for computing correlation functions of operators in the 1/2 BPS sector from holography using the so-called LLM geometries \cite{Lin:2004nb}. A detailed, pedagogical introduction can be found in \cite{Skenderis:2007yb}, and application to HHL correlators with giant gravitons in \cite{Caputa:2012yj} (see also \cite{Anempodistov:2025maj} for recent applications). Intuitively, in this approach to HHL correaltors, we can construct an explicit back-reacted geometry corresponding to heavy $\chi_R(Z)$, and compute one-point functions of the light probe $\Tr(Z^J)$ in this background.

More precisely, the LLM approach to correlators is based on the second-quantized fermion field
\bea
  \Psi (z,z^*,t)=\sum_{l=0}^\infty \hat{C}_l e^{-i(l+1)t}\Phi_l(z,z^*)\;,
\eea
with oscillators obeying the anti-commutators $\{ \hat{C}_l,\hat{C}_m^\dagger\} =\delta_{lm}$, and the mode functions are the orthonormal wave functions of the lowest Landau level
\bea
  \Phi_l(z,z^*)=\sqrt{2^{l+1}\over\pi\, l!}z^l e^{-zz^*}\;.
\eea
By construction, the setup consists of $N$ fermions so the second quantized field obeys
\bea
  \int dzdz^* \Psi^\dagger (z,z^*,t)\Psi (z,z^*,t) =\sum_{l=0}^\infty \hat{C}_l^\dagger \hat{C}_l = N\;.\label{NormPsi}
\eea
The Schur polynomials $\chi_R(Z)$ correspond to energy eigenstates of the $N$ fermion system and their energies are
\be
E_i=r_i+N-i+1 \, \qquad i=1,...,N\,,
\ee
where $r_i$ is the number of boxes in the $i$-th row of the Young diagram $R$. 
This way, we can associate with $\chi_R(Z)$ a quantum state
\bea
  |\chi_R\rangle\equiv \hat{C}_{N-1+r_1}^\dagger\hat{C}_{N-2+r_2}^\dagger \cdots \hat{C}_{1+r_{N-1}}^\dagger\hat{C}_{r_N}^\dagger |0\rangle\;.
\eea
Next, given this state, we define
$$
  U_{lm}=\langle \chi_R|\hat{C}_l^\dagger \hat{C}_m|\chi_R\rangle\,,
$$
from which we obtain the density function that determines the LLM geometry \cite{Lin:2004nb} as
\bea
  \rho={1\over 2}\sum_{l,m} (z^*)^l z^m \sqrt{2^{2+l+m}\over \pi^2\, l!\, m!}e^{-2zz^*}U_{lm}\;.
  \label{LLMdensity}
\eea
As a consequence of normalization \eqref{NormPsi}, 
these densities satisfy
\bea
  \int_0^{2\pi} d\phi\int_0^\infty \, r dr\, \rho = {N\over 2}\;.
  \label{Normedrho}
\eea
Finally, the computation of one point function amounts to the integral \cite{Skenderis:2007yb}
\bea
  \left< G\left|{{\rm Tr} (Z^J)\over \sqrt{JN^J}}\right|G \right> =  {N\over \sqrt{J}}\int r^{J}\rho\, e^{iJ\phi}\, r dr\, d\phi\;.
  \label{holone}
\eea
For our sub-determinants we must consider a quantum state corresponding to a linear combination of the two operators $\chi_{(1^k)}(Z)$ and  $\chi_{(1^{k-J})}(Z)$ \cite{Caputa:2012yj}
\bea
    |G\rangle =|\chi_{(1^k)}\rangle +|\chi_{(1^{k-J})}\rangle\;,
    \label{SG}
\eea
or in terms of fermionic oscillators 
\bea
  |G\rangle =(\hat{C}^\dagger_{N-k+J-1}+\hat{C}^{\dagger}_{N-k-1})\prod_{i=0,\ne N-k-1,\ne N-k+J-1}^N\hat{C}^\dagger_{i}|0\rangle\,.
\eea
Then we can compute the matrix $U_{lm}$ and extract the density 
\bea
  \rho = {1\over \pi}e^{-iJ\phi}{(2z^*z)^{N-k-1+{J\over 2}}\over \, \sqrt{(N-k-1)!(N-k+J-1)!}}e^{-2zz^*}\,,\qquad z\bar{z}=\frac{N}{2}r^2\,,
\eea
such that the final answer for the correlator becomes
\bea
  \left< G\left|{{\rm Tr} (Z^J)\over \sqrt{JN^J}}\right|G \right> = \sqrt{\frac{\Gamma(N-k+J)}{JN^J\Gamma(N-k)}}\;.\label{finalcorrelator}
\eea
Notice that the LLM computation almost reproduces the CFT answer \eqref{GenForm} with $\Gamma(\cdot)$ functions (and large-N approximation of the chiral-primary norm).

Finally, analogously to the CFT computation \eqref{3ptAHHL}, we can then extract the HHL limit \eqref{HHLLimit} 
\bea
  \left< G\left|{{\rm Tr} (Z^J)\over \sqrt{JN^J}}\right|G \right> = {1\over\sqrt{J}} \left(1-{k\over N}\right)^{J\over 2}\;,
\eea
recovering the CFT result.
\section{Evanescent States and the Giant Graviton Expansion}\label{sec:EvanescentStates}
In this section we analytically continue conformal field theories with gauge group $U(N)$, where $N$ is usually a positive integer, to arbitrary complex rank. Continuations of this type are familiar from the $\epsilon$-expansion in conformal field theory, where the spacetime dimension $d$ is analytically continued away from integer values. This analogy provides useful intuition: at the Wilson--Fisher fixed point, for example, one encounters {\it evanescent states} that exist only for non-integer $d$ and vanish identically when $d$ is specialized to an integer. In Section~\ref{evanescent} we identify completely analogous states that arise when $N$ is analytically continued. A simple example is provided by a giant graviton carrying angular momentum greater than the maximal value allowed at finite $N$. For integer $N$ such states vanish, and their disappearance is nothing but a trace relation. In this concrete sense, evanescent states are directly tied to trace relations, and it is natural to conjecture that they correspond to the ghost brane states of the bulk dual~\cite{Gaiotto:2021xce,Lee:2023iil,Lee:2024hef}. With this identification in hand, we derive a giant graviton-like expansion in section~\ref{ggexpansion}. In section~\ref{invggexpansion} we show that our expansion is mathematically identical to the giant graviton expansion of~\cite{Gaiotto:2021xce}.
\subsection{Evanescent states}\label{evanescent}
The Wilson-Fischer CFT exhibits a fascinating and subtle spectrum of {\it evanescent states}--operators\footnote{These should not be confused with the evanescent operators described for example in \cite{Rey:2014dpa,Jevicki:2015sla,Jevicki:2015pza,Amado:2017kgr,Banerjee:2019iwd,Engelsoy:2021fbk}, which signal the presence of a horizon in the holographic dual to a CFT. These modes are present for integer values of $N$.} that exist only when the spacetime dimension $d$ is not an integer. These states correspond to operators that vanish identically at particular integer dimensions. They involve totally antisymmetric combinations of indices that are impossible when $d$ is too small, but are non-zero when $d$ is analytically continued away from integers. In the scalar $\phi^4$ theory, for example, evanescent operators appear only at relatively large dimensions (e.g. $\Delta \approx 23$ at $d=4-\epsilon$, and can have {\it negative norm} under the continued Gram (inner-product) matrix--signaling a breakdown of unitarity in non-integer dimensions. Further, the operators corresponding to negative norm states can not be dropped because they appear in the operator product expansion (OPE) of two physical operators. A key reference for this background material, which was an important motivation for our study, is \cite{Hogervorst:2015akt}.

It is worth quickly reviewing how in conformal field theory we define the theory in non-integer spacetime dimensions 
$d$. We treat $d$ as a continuous complex parameter in all the integrals, tensor identities, and gamma functions that appear in perturbation theory. We define correlation functions, beta functions, anomalous dimensions, etc. by dimensional regularization i.e. we write integrals in $d$ dimensions and evaluate them for general complex $d$. Loop integrals are defined in $d$ dimensions as
\begin{equation}
\int {d^d p\over (2\pi)^d}{1\over (p^2+m^2)^\alpha}\,.
\end{equation}
These are evaluated by switching to spherical coordinates in $d$ dimensions, using
\begin{equation}
\int d^d p\cdots ={2\pi^{d\over 2}\over\Gamma(d/2)}\int_0^\infty dp \, p^{d-1}\cdots\,.
\end{equation}
This formula is valid for any positive integer $d$. After evaluating the integral, we typically get results involving $\Gamma$-functions of $d$. For example
\begin{eqnarray}
\int {d^d p\over (2\pi)^d}{1\over p^2+m^2}&=&{1\over (4\pi)^{d\over 2}}\Gamma\left(1-{d\over 2}\right) (m^2)^{{d\over 2}-1}\,.
\end{eqnarray}
Thanks to the appearance of the $\Gamma(\cdot)$ function, this is a perfectly analytic function of $d$, so we can continue it to arbitrary values of $d$.

Now consider $U(N)$ ${\cal N}=4$ super Yang--Mills theory. Correlation functions of trace operators yield polynomials in $N$. Since polynomials are entire functions of finite order, they admit a unique analytic continuation. In sectors such as the ${1\over 2}$-BPS sector, the Schur polynomials provide a natural basis for local operators. This basis automatically implements the trace relations by restricting to Schur polynomials labelled by Young diagrams with at most $N$ rows. Moreover, as shown in the previous section, the $N$-dependence of correlation functions of Schur polynomials is expressed through ratios of $\Gamma$-functions, making the continuation to arbitrary (even non-integer or complex) values of $N$ straightforward.

There are also evanescent states in $U(N)$ ${\cal N}=4$ super Yang-Mills theory, when we continue $N$ to non-integer values. For example, consider the Schur polynomial two point function, with each operator labelled by a Young diagram that is a single column with $n$ boxes
\begin{eqnarray}
\langle\chi_{(1^n)}(Z(x_1))^\dagger \chi_{(1^n)}(Z(x_2))\rangle &=& {N!\over (N-n)!} {1\over |x_1-x_2|^{2n}}
={\Gamma(N+1)\over \Gamma(N-n+1)}{1\over |x_1-x_2|^{2n}}\,.\nn\\
\end{eqnarray}
To clarify the discussion that follows, when $N\in\mathbb{Z}$ we use the notation $N$ and when we have continued $N$ to obtain a general $N\in\mathbb{C}$ we use the notation $N^*$, i.e. we are defining the correlators of the $U(N^*)$ theory by analytic continuation of the correlators of the $U(N)$ theory. To illustrate the existence of evanescent states, choose $N^*=2.5$ and $n=4$, to obtain
\begin{eqnarray}
\langle\chi_{(1^4)}(Z(x_1))\chi_{(1^4)}(Z(x_2))\rangle&=&N^*(N^*-1)(N^*-2)(N^*-3){1\over |x_1-x_2|^8}\cr\cr
&=&-{15\over 16}{1\over |x_1-x_2|^8}<0\,,
\end{eqnarray}
so that this is indeed an evanescent state. Here we are using the fact that the two point function defines a norm on the Hilbert space of the radially quantized CFT. The fact that the above two point function is negative implies that the state corresponding to this operator has a negative norm. Further, just like in the case of the Wilson-Fisher CFT, these operators vanish when $N^*$ is set to 2 or 3 and these evanescent states are generated in the OPE of two physical states. To see this last point, first note that $\chi_{(1^2)}(x)$ is a physical state, it is not evanescent, as it does not vanish at $N^*=2,3$. Indeed, consider the two point function, which is
\begin{eqnarray}
\langle \chi_{(1^2)}(x_1)^\dagger\chi_{(1^2)}(x_2)\rangle&=&N^*(N^*-1){1\over |x_1-x_2|^4}
={15\over 4}{1\over |x_1-x_2|^4}\,.
\end{eqnarray}
Now consider a product of two such operators, which gives\footnote{The labels of the Schur polynomials list row lengths so that $(2^2)={\tiny\yng(2,2)}$ and $(21^2)=\tiny\yng(2,1,1)$.}
\begin{equation}
 \chi_{(1^2)}(x_1)\chi_{(1^2)}(x_2)=\chi_{(1^4)}\left({x_1+x_2\over 2}\right)+\chi_{(21^2)}\left({x_1+x_2\over 2}\right)+\chi_{(2^2)}\left({x_1+x_2\over 2}\right)+\cdots\,.
\end{equation}
We chose to take the OPE around the midpoint between the two operators and $\cdots$ reflects terms corresponding to descendants which necessarily include derivatives. The identity block does not contribute because even though the two operators appearing have the same dimension, they have the same (not opposite) $R$-charge so their two point function vanishes.  Notice that the first term that appears is the evanescent operator, so it is indeed generated in the OPE of two non-evanescent operators.

Just as for the Wilson-Fisher CFT, not all evanescent operators correspond to states that have negative norm. For example, when $N^*=2.5$, $\chi_{(1^5)}(x)$ is evanescent but 
\begin{equation}
    \langle\chi_{(1^5)}(x_1)^\dagger\chi_{(1^5)}(x_2)\rangle={45\over 32}{1\over |x_1-x_2|^{10}}>0\,.
\end{equation}

To summarize, after continuing $N$ to $N^*$ we find that the spectrum of the theory includes negative norm states, signalling that the $U(N^*)$ theory is not unitary. These negative norm states are Schur polynomial states labelled by Young diagrams with more rows than the continuous value of $N^*$. These negative norm states vanish when $N^*$ is increased to an integer value and their vanishing is a consequence of trace relations. This strongly suggests that we should interpret the evanescent states exhibited above as the CFT state dual to the bulk ghost brane states that appears in the analysis of~\cite{Gaiotto:2021xce,Lee:2023iil,Lee:2024hef}. 

The CFT computation of correlation functions involving the evanescent states simply entails performing an analytic continuation in $N$. This is straightforward since, as remarked above, the relevant correlators depend on $N$ through the $\Gamma$ function. Since correlators can be interpreted as transition amplitudes of the radially quantized CFT, this opens up the possibility that we can compute transition amplitudes between physical and ghost brane states. We will develop this idea in Section \ref{sec:TransitionAmplitudes} below. Another consequence of this identification is that it allows us to find a new formula for the $\frac{1}{2}$-BPS index, which is closely related to the giant graviton expansion. We develop this formula next.

\subsection{$\frac{1}{2}$-BPS index}\label{ggexpansion}
Every $\frac{1}{2}$-BPS multiplet of ${\cal N}=4$ super Yang-Mills theory has a unique operator constructed entirely from a complex adjoint matrix $Z$. The dimension of this complex scalar equals its $R$ charge equals 1. Counting BPS multiplets is then reduced to the problem of counting the gauge invariant operators that can be constructed from $Z$. As a consequence, the $\frac{1}{2}$-BPS index has a simple combinatorial interpretation: it counts the gauge-invariant operators built from the adjoint BPS letter corresponding to $Z$. Equivalently, the $\frac{1}{2}$-BPS index can be evaluated by computing the partition function of a matrix oscillator, with the ground state energy subtracted off and the spacing between energy levels fixed to 1.  In what follows we take $x=e^{-\beta}$.

Consider the computation of the partition function for a one matrix model. There are two important features of the $U(N^*)$ theory that differ from the $U(N)$ theory:
\begin{itemize}
    \item When $N^*\notin\mathbb{Z}$ as we have explained above there are no trace relations. Trace relations imply that not all operators are independent, so they reduce the value of the $\frac{1}{2}$-BPS index.
    \item There are the new evanescent states that must be accounted for. Note that these contributions vanish in the $U(N)$ theory and are not included in the index of the integer $N$ theory.
\end{itemize}
The first point above implies that the complete space of gauge invariant operators is freely generated by all single trace operators with no cut off imposed on the number of matrices in the trace. Thus, the partition function is simply counting partitions and we immediately obtain
\begin{equation}
    Z_{\rm phys}(x)=\frac{1}{\prod_{i=1}^\infty(1-x^i)}\,.
\end{equation}
The subscript ``phys'' above indicates that we have computed the contribution from the free ring, but have not yet taken effects due to the evanescent states into account. Let us now turn to the contributions of the evanescent states. As emphasized in~\cite{Lee:2023iil}, ghost-brane states admit a natural interpretation: they correspond to giant gravitons that compensate for null states arising from trace relations. In this picture, the open strings attached to the bulk D-branes are not dual to physical states of the gauge theory; rather, they systematically cancel the redundancies among states that appear at finite (integer) $N$. Since we are proposing that evanescent states are dual to ghost-brane states, we adopt this interpretation here. Accordingly, the contribution of evanescent states must be subtracted. An evanescent state with $k$ columns and $\lceil N^*\rceil+1$ rows is then naturally interpreted as a bound state of $k$ evanescent giant gravitons. Denoting the partition function of such a state, including all closed- and open-string excitations, by $Z_k(x)$, our proposal is that the full partition function $Z(x)$ takes the form
\begin{equation}
    Z(x)=Z_{\rm phys}(x)-\sum_{k=1}^\infty Z_k(x)\,.
\end{equation}
To proceed, we will now evaluate the partition function $Z_k(x)$ which captures the contribution from $k$-evanescent giant gravitons. Each state is most transparently labelled with a Young diagram as shown below:
$$
{\footnotesize \young(***{\,}{\,}{\,}{\,}{\,}{\,}{\,}{\,}{\,}{\,},***{\,}{\,}{\,}{\,}{\,}{\,}{\,}{\,},***{\,}{\,}{\,}{\,}{\,}{\,},***{\,}{\,}{\,}{\,},***{\,}{\,},***,{\,}{\,}{\,},{\,}{\,}{\,},{\,}{\,},{\,}{\,},{\,},{\,},{\,})}
$$

\noindent
The starred boxes represent the evanescent brane bound state, with the number of columns reflecting the number of branes in the bound state. The bound state shown is a system of three branes. The number of rows is given by $\lceil N^*\rceil +1$. The example shown has $N^*=4.5$. Closed string excitations are stacked in columns to the right of the brane state with a maximum height of $\lceil N^*\rceil$. We restrict the maximum height so that the allowed excitations of a $k$-evanescent brane system can not produce a $k+1$-evanescent brane system. Thus, the closed string excitation is specified by $N=\lceil N^*\rceil$ integers $\{n_1,n_2,\cdots,n_N\}$ where $n_l$ counts the number of columns of length $l$. Open string excitations are stacked as additional rows below the brane state, with a maximum length of $k$. This reflects the fact that the world volume theory of $k$ giant gravitons is a $U(k)$ gauge theory and because $k$ is an integer, trace relations ensure that the invariants constructed in this theory are polynomials of single trace operators with no more than $k$ matrices in a trace. The open string excitation is thus labelled by $k$ integers $\{m_1,m_2,\cdots,m_k\}$, where $m_l$ counts the number of rows of length $l$. Thus, each evanescent brane state is labelled by $N+k$ integers $\{n_1,\cdots,n_N,m_1,\cdots,m_k\}$ with the only condition being that each integer is not negative. For the state shown above we have $m_1=3$, $m_2=m_3=2$ and $n_1=n_2=n_3=n_4=n_5=2$. The energy of the state is given by total the number of boxes (i.e. starred boxes are counted) in the Young diagram. The partition function which sums the states of the bound state of $k$-evanescent giant gravitons is thus given by
\begin{eqnarray}
    Z_k(x)&=&x^{\lceil N^*\rceil k+k}\sum_{m_1=1}^\infty\cdots \sum_{m_k=1}^\infty
     (x)^{m_1}(x^2)^{m_2}\cdots (x^k)^{m_k}
     \sum_{n_1=1}^\infty\cdots \sum_{n_N=1}^\infty
     (x)^{n_1}(x^2)^{n_2}\cdots (x^N)^{n_N}\cr\cr
     &=&\frac{x^{\lceil N^*\rceil k+k}}{\prod_{i=1}^k(1-x^i)\prod_{j=1}^{\lceil N^*\rceil}(1-x^j)}\,.\label{EBraneCount}
\end{eqnarray}
This formula is demonstrated by giving a one-to-one correspondence between the terms summed in the partition function and the states of the evanescent branes which are labelled by Young diagrams. To see the bijection note that there is a unique term on the RHS of the first equality in (\ref{EBraneCount}) for a given set of the $N+k$ integers $\{n_1,\cdots,n_N,m_1,\cdots,m_k\}$, with the only condition being that each integer is not negative. This demonstrates the bijection because this set of integers, with these conditions, specify a unique evanescent brane state, and every possible evanescent brane is given by a collection of integers that is within the range of the above sum.

As a check of this proposal, we know that when $N=\lceil N^*\rceil$ that the partition function is given by
\begin{equation}
    Z(x)=\frac{1}{\prod_{i=1}^N(1-x^i)}\,,
\end{equation}
so that we predict the giant graviton-like expansion
\begin{equation}
    \frac{1}{\prod_{i=1}^N(1-x^i)}=\frac{1}{\prod_{i=1}^\infty(1-x^i)}-\sum_{k=1}^\infty\frac{x^{N k+k}}{\prod_{i=1}^k(1-x^i)\prod_{j=1}^{N}(1-x^j)}\,.\label{tracerelbranes}
\end{equation}
It is rather simple to verify that this identity is true.

To summarize, continuing $N\in\mathbb{Z}$ to $N^*\in\mathbb{R}$, the trace relations are removed and there are additional contributions coming from evanescent states. The evanescent states can have negative norm and they imply that after continuation the theory is no longer unitary. We have proposed that these new states should be identified with the ghost-brane states introduced in \cite{Lee:2023iil}. According to \cite{Lee:2023iil} these states (and their excitations) are not physical states, but rather they cancel redundancies that should have been cancelled by the trace relations. Our formula (\ref{tracerelbranes}) confirms that this is exactly what happens. From this point on we will no longer indicate the analytically continued $N$ as $N^*$.

\subsection{Giant Graviton Expansion}\label{invggexpansion}
Although our formula (\ref{tracerelbranes}) does not exactly match the giant graviton expansion of \cite{Gaiotto:2021xce} in all of its details, there is clearly a complete parallel between the description in terms of the $U(N^*)$ gauge theory and the dual bulk description developed in~\cite{Gaiotto:2021xce,Lee:2023iil,Lee:2024hef}. In this section we will develop the precise connection between the two formulas.

To motivate the giant graviton expansion, note that we can write the finite $N$ ${1\over 2}$-BPS index as follows
\be
Z_N(q)=\frac{1}{\prod^N_{n=1}(1-q^n)}=\frac{1}{\prod^\infty_{n=1}(1-q^n)}\prod^N_{n=1}(1-q^{N+n})\,.
\ee
The product on the right can be expanded to obtain
\be
\frac{1}{\prod^N_{n=1}(1-q^n)}=\frac{1}{\prod^\infty_{n=1}(1-q^n)}\sum^\infty_{k=0}(-1)^kq^{kN}\frac{q^{k(k+1)/2}}{\prod^k_{l=1}(1-q^l)}\,.\label{ggExp}
\ee
Using the identification
\be
(-1)^k\frac{q^{k(k+1)/2}}{\prod^k_{l=1}(1-q^l)}=\frac{1}{\prod^k_{n=1}(1-q^{-n})}=Z_k(q^{-1})\,,
\ee
we obtain the giant graviton expansion~\cite{Gaiotto:2021xce}
\be
Z_N(q)=Z_\infty(q)+\sum^\infty_{k=1}q^{kN}Z_k(q^{-1})\,.
\ee

These formulas have a number of interesting features. The overall factor on the right hand side of (\ref{ggExp}) is (up to the factor of $q^{1/24}$) the Dedekind eta function
\be
\eta(\tau)=q^{1/24}\prod^\infty_{n=1}(1-q^n)\,,\qquad q=e^{2\pi i\tau}\,.
\ee
Recall that under modular S: $\tau\to -1/\tau$, and T: $\tau\to \tau+1$, transformations we have
\be
\eta(-1/\tau) = \sqrt{-i\tau}\, \eta(\tau)\,,\qquad \eta(\tau+1)=e^{\frac{\pi i}{12}}\eta(\tau)\,.
\ee
If we don't care about modularity then most of the story can be discussed in terms of the Euler function
\be
\phi(q)=q^{-1/24}\eta(\tau)=\prod^\infty_{n=1}(1-q^n)\,.
\ee
A natural finite $N$ regularisation of the eta function (which gives up modularity) is
\be
\eta_N\equiv q^{1/24}\prod^N_{n=1}(1-q^n)\,.
\ee
This is naturally extended to non-integer $N$ using the q-Pochhammer symbols
\be
(q;q)_x=\prod^x_{n=1}(1-q^n)\,.
\ee
The q-Pochhammer symbols satisfy identities that may have interesting consequences for giant-graviton like expansions. For example, our identity (\ref{tracerelbranes}) (setting $x\to q$)
\begin{equation}
    \frac{1}{\prod_{i=1}^N(1-q^i)}=\frac{1}{\prod_{i=1}^\infty(1-q^i)}-\sum_{k=1}^\infty\frac{q^{k(N+1)}}{\prod_{i=1}^k(1-q^i)\prod_{j=1}^{N}(1-q^j)}\,,
\end{equation}
has an interpretation as an identity for q-Pochhammer symbols
\be
\frac{1}{(q;q)_N}=\frac{1}{(q;q)_\infty}-\frac{1}{(q;q)_N}\sum_{k=1}^\infty\frac{q^{k(N+1)}}{(q;q)_k}\,,
\ee
or equivalently
\be
\frac{(q;q)_N}{(q;q)_\infty}=1+\sum_{k=1}^\infty\frac{q^{k(N+1)}}{(q;q)_k}\,.\label{NNiceIddent}
\ee
This identity has an unambiguous continuation to non-integer $N$. The expression on the right hand side above can be summed to
\be
1+\sum_{k=1}^\infty\frac{q^{k(N+1)}}{(q;q)_k}=\frac{1}{(q^{N+1};q)_\infty}=\frac{1}{\prod^\infty_{n=0}(1-q^{(N+1)}q^{n})}\,,
\ee
while the expression on the left hand side is
\be
\frac{(q;q)_N}{(q;q)_\infty}=\frac{\prod^N_{n=1}(1-q^n)}{\prod^N_{n=1}(1-q^n)\prod^\infty_{n=N+1}(1-q^n)}=\frac{1}{\prod^\infty_{n=0}(1-q^{N+1}q^n)}\,,
\ee
which proves (\ref{NNiceIddent}) as long as $|q|<1$ which is indeed the case. Using the relation
\be
(q;q)_k=(-1)^kq^{k(k+1)/2}(q^{-1};q^{-1})_k\,,
\ee
we can write this expression as
\be
\frac{(q;q)_N}{(q;q)_\infty}=1+\sum_{k=1}^\infty q^{kN}\frac{(-1)^kq^{-\frac{k(k-1)}{2}}}{(q^{-1};q^{-1})_k}\,.
\ee
In fact we can also write the giant graviton expansion \eqref{ggExp} as
\be
\frac{(q;q)_\infty}{(q;q)_N}=1+\sum^\infty_{k=1}q^{kN}\frac{1}{(q^{-1};q^{-1})_k}\,.
\ee
Notice that both the giant graviton expansion and our formula (\ref{tracerelbranes}) are simply identities for q-Pochhammer symbols. The giant graviton expansion gives a formula for $\frac{(q;q)_\infty}{(q;q)_N}$, while our expansion (\ref{tracerelbranes}) gives a formula for $\frac{(q;q)_N}{(q;q)_\infty}$. The equivalence between our expansion (\ref{tracerelbranes}) and the giant graviton expansion follows immediately from
\be
\frac{(q;q)_N}{(q;q)_\infty}\frac{(q;q)_\infty}{(q;q)_N}=1=\left(\sum_{k=0}^\infty q^{kN}\frac{(-1)^kq^{-\frac{k(k-1)}{2}}}{(q^{-1};q^{-1})_k}\right)\left(\sum^\infty_{k=0}q^{kN}\frac{1}{(q^{-1};q^{-1})_k}\right)
\ee
which is easily verified. This gives a precise relation between the giant graviton expansion of \cite{Gaiotto:2021xce} and identity (\ref{tracerelbranes}).
\section{Transition Amplitudes}\label{sec:TransitionAmplitudes}
Our analysis reveals new evanescent states, which we identify as the states dual to giant gravitons with angular momentum exceeding the maximal giant bound set by $N$. A concrete description of such states directly in the bulk has been developed in an interesting series of papers ~\cite{Gaiotto:2021xce,Lee:2023iil,Lee:2024hef}. These states are naturally associated with trace relations. In this Section, we develop some of the properties of these trace relation states: we explain how to evaluate transition amplitudes between states arising from trace relations and ordinary physical giant graviton states. This analysis reveals a natural connection between the work of~\cite{Gaiotto:2021xce,Lee:2023iil,Lee:2024hef} and our proposal of analytic continuation in $N$: the transition amplitudes precisely match ordinary CFT amplitudes analytically continued in $N$.

The next section shows that analytic continuation in $N$ can be understood mathematically as working in the free ring generated by all traces, with no relations imposed. Reviewing key results of~\cite{Gaiotto:2021xce,Lee:2023iil,Lee:2024hef}, we then explain how trace relations can be imposed systematically by introducing ghosts to represent the states associated with them, together with a fermionic charge and the associated Koszul complex. With this framework in place, in \ref{transitions}  we compute transition amplitudes and find that they match precisely with ordinary CFT amplitudes analytically continued in $N$.
\subsection{Free half-BPS ring $R$}
Our answers for correlators are obtained by analytic continuation, to non-integer $N$, of formulas obtained in the CFT.  This is equivalent to working in the holomorphic half–BPS sector, with the ring
\bea
R \;\equiv\; \mathbb{C}[p_1,p_2,p_3,\ldots], \qquad p_m(Z) \equiv \Tr Z^m,
\eea
where there are no relations among the single traces $p_m(Z)$. $R$ is just the free ring of all polynomials in the complete set of single trace operators and consequently, $R$ includes the complete set of all gauge invariant operators. Correlation functions provide a way to map pairs of elements $O_A(Z), O_B(Z)\in R$ to numbers $\langle O_A(Z^\dagger)O_B(Z)\rangle$. The rule to evaluate correlators, is to use Wick's theorem with the basic two point function given by the usual matrix model Wick contractions, which are normalized as
\bea
\label{WickRule}
\langle Z_{ij}\, Z^\dagger_{kl}\rangle \,\,=\,\, \delta_{il}\,\delta_{jk}\,,
\qquad
\langle Z_{ij}\, Z_{kl}\rangle\,\,=\,\,\langle Z^\dagger_{ij}\, Z^\dagger_{kl}\rangle=0\,.
\eea
After using these Wick contractions the correlator follows by summing over matrix indices, using the usual rule for the Kronecker delta function. The only way that $N$ appears is through the sum
\bea
\delta_{ii}&=&N^*\,.
\eea
When working in the free ring $R$ we are completely agnostic about the value of $N$ and simply leave it as an undetermined parameter. In particular, we do not assume it is a positive integer and it is for this reason that we denote it as $N^*$. When $N^*$ is not a positive integer, there are no trace relations constraining the theory.

The point of view we take in this section will prove useful below. In particular, we will be able to reduce the computation of transition amplitudes between unphysical and physical giants to the computation of a correlator inside $R$, carried out using the above rules.

\subsection{The physical half-BPS ring at finite integer $N$}
To recover the physical space of half-BPS operators from $R$ we need to set $N^*$ equal to a positive integer and impose the resulting trace relations. 
\paragraph{Finite-$N$ trace ideal.}
At finite positive integer $N$, traces satisfy polynomial relations. Let $I_N\subset R$ be the (two–sided, graded) ideal generated by the independent trace relations. The trace relations express $p_{N+1+a}(Z)$ in terms of $\{p_1(Z),\dots,p_{N}(Z)\}$, together with their polynomial consequences, for any integer $a\ge 0$. A nice way to generate the complete set of trace relations is from the Schur polynomials. Let $\chi_{(p^k)}(Z)$ be the Schur polynomial labelled by a Young diagram given by $p$ columns and $k$ rows. Then $I_N$ is generated by $\chi_{(1^{N+1+a})}(Z)$. The physical finite-$N$ ring is the quotient
\bea
H_N \;\cong\; R/I_N\,.\label{quotientwewant}
\eea

Following ~\cite{Lee:2023iil}, we can relate $H_N$ and $R$ by introducing states that correspond to giants with angular momentum $k>N$. These are ghost states described by making use of an anti-commuting field, $\eta_X$ where we will spell out the subscript below. Further, in this description we will grade by ghost number. These extra states keep track of the relations that must be imposed on $R$ to obtain $H_N$.
\paragraph{Ghosts and graded modules.}
Introduce fermionic generators (``ghosts'') $\eta_{-(N+1+a)}$ for $a\in\mathbb{Z}_{\ge 0}$, one for each generator of $I_N$. Denote the $\C$–vector space spanned by these ghosts as
\bea
E \;=\; \mathrm{span}_\C\{\eta_{-(N+1+a)}\}_{a\ge 0}\,,
\eea
and grade by ghost number
\bea
\deg(\eta_X)=+1\,.
\eea
We also make use of the exterior powers $\Lambda^k E$ to define the graded $R$-modules
\bea
V_k^N \;=\; R \otimes \Lambda^k E\,,\qquad (k=0,1,2,\ldots)\,,
\eea
with {\it ghost number} $k$. We introduce a fermionic charge $Q_b$, that relates the fermionic ghosts to the trace relations as follows:

\paragraph{Definition of $Q_b$} $\mathbf{p}=\{\chi_{(1^{N+1+a})}(Z)\}_{a\ge 0}$ is our generating set of $I_N$. Define $Q_b$ by
\bea
\label{eq:Q-on-ghost}
Q_b\big(\eta_{-(N+1+a)}\big) \;=\; \chi_{(1^{N+1+a})}(Z)\,,\qquad Q_b\big(p_m(Z)\big)=0\,,
\eea
and extend $Q_b$ as a graded derivation ($W\in R$)
\bea
\label{eq:graded-Leibniz}
Q_b\big( W \otimes \eta_{i_1}\!\wedge\!\cdots\!\wedge\!\eta_{i_k} \big)=
\sum_{r=1}^k (-1)^{r-1} \big( W\, Q_b\eta_{i_r} \big)\otimes
\eta_{i_1}\!\wedge\!\cdots\widehat{\eta_{i_r}}\cdots\!\wedge\!\eta_{i_k}\,,
\eea
with $Q_b(W)=0$ and $\widehat{\eta_{i_r}}$ indicates that $\eta_{i_r}$ is omitted. This makes $Q_b:V_{k}^N\to V_{k-1}^N$ of degree $-1$ and it ensures that $Q_b^2=0$.

The modules introduced above can be organized into a {\it chain complex}~\cite{Lee:2023iil}. A chain complex is an algebraic structure given by a sequence of modules and a sequence of homomorphisms between consecutive modules such that the image of each homomorphism is contained in the kernel of the next~\cite{mccleary2001user}. So a chain complex is not an exact sequence in general. In fact, the homology of a chain complex measures the failure of the chain complex to be exact.

\paragraph{The complex:} Organize the $V_k^N$ into a chain complex
\begin{equation}
\label{eq:complex}
\cdots \xrightarrow{\,Q_b\,} V_3^N \xrightarrow{\,Q_b\,} V_2^N \xrightarrow{\,Q_b\,} V_1^N \xrightarrow{\,Q_b\,} V_0^N=R \xrightarrow{\,\pi\,} H_N \to 0,
\end{equation}
where $\pi$ is the natural projection $R\to R/I_N$.

It is easy to see how the chain complex is constructed at low ghost number. For $a,b,c\in\mathbb{Z}_{\ge 0}$ and $W\in R$ we have
\bea
&&\text{($k=1$)} \qquad Q_b\big(W(Z)\,\eta_{-(N+1+a)}\big)\,\,=\,\,W(Z)\,\chi_{a}(Z)\cr\cr\,
&&\text{($k=2$)} \qquad Q_b\big(W(Z)\,\eta_{-(N+1+a)}\!\wedge\!\eta_{-(N+1+b)}\big)
\,\,=\,\, W(Z)\,\chi_{a}(Z)\,\eta_{-(N+1+b)}\cr\cr
&&\qquad\qquad\qquad\qquad\qquad\qquad\qquad\qquad\qquad\qquad
-W(Z)\,\chi_{b}(Z)\,\eta_{-(N+1+a)}\cr\cr\,
&&\text{($k=3$)} \qquad Q_b\big(W(Z)\,\eta_{a}\!\wedge\!\eta_{b}\!\wedge\!\eta_{c}\big)
= W(Z)\Big(\chi_a(Z)\,\eta_{b}\!\wedge\!\eta_{c}- \chi_b(Z)\,\eta_{a}\!\wedge\!\eta_{c}+\chi_c(Z)\,\eta_{a}\!\wedge\!\eta_{b}\Big)\nonumber
\eea
where we have set $\chi_a(Z)\equiv \chi_{(1^{N+1+a})}(Z)$ and $\eta_a\equiv \eta_{-(N+1+a)}$ to ease the notation. This construction coincides with a Koszul complex. The Koszul complex starts from a commutative ring $A$, and an $A$-linear map $s: A^r\to A$. Here $A^r$ stands for the free $A$-module of rank $r$. Concretely, $A^r$ is the direct sum of $r$ copies of $A$
\bea
A^r&=&A\oplus A\oplus A\oplus\cdots\oplus A\,.
\eea
As a set, it consists of all $r$-tuples $(a_1,a_2,\cdots,a_r)$ with each $a_i\in A$. As a module, addition and scalar multiplication are defined component-wise
\bea
(a_1,a_2,\cdots,a_r)+(b_1,b_2,\cdots,b_r)&=&(a_1+b_1,a_2+b_2,\cdots,a_r+b_r)\,,
\eea
\bea
c (a_1,a_2,\cdots,a_r)=(ca_1,ca_2,\cdots,ca_r)\,,
\eea
for $c\in A$. If $A$ is a field (for example $A=\mathbb{C}$), then $A^r$ is just the usual vector space of dimension $r$. The Koszul complex $K_s$ is
\bea
\bigwedge^{r}A^r\ \to \ \bigwedge ^{r-1}A^r\ \to \ \cdots \ \to \ \bigwedge ^{1}A^r\ \to \ \bigwedge ^{0}A^r\simeq A\,,
\eea
where $\bigwedge^{l}A$ is the antisymmetric (wedge) product and where the maps are defined explicitly as follows
\bea
\alpha _{1}\wedge \cdots \wedge \alpha _{k}\ \mapsto \ \sum _{i=1}^{k}(-1)^{i+1}s(\alpha _{i})\ \alpha _{1}\wedge \cdots \wedge {\hat {\alpha }}_{i}\wedge \cdots \wedge \alpha _{k}\,,
\eea
where again $\hat {\ }$ means the term is omitted and $\wedge$ is the wedge product. Comparing to the definition of the action of $Q_b$, this clearly matches the complex constructed using $Q_b$ and the ghost modules above.

The homology of the Koszul complex tells us when a set of elements of a (local) ring is an $M$-regular sequence, i.e. it exposes basic facts about the depth of a module. In certain circumstances the complex is the complex of syzygies \cite{eisenbud2013commutative} so it tells us the relations between generators of a module, the relations between these relations, etc. If we treat the $\{\chi_{(1^{N+1+a})}(Z)\}$ as a regular sequence, the complex \eqref{eq:complex} is the Koszul complex for $I_N$, with
\bea
H^0(Q_b)\;\cong\; R/I_N \;=\; H_N,\qquad H^{k>0}(Q_b)\;=\;0\,.
\eea

\subsection{Transition matrix elements}\label{transitions}
To compute a transition amplitude, we need an inner product. To define this inner product it is useful to introduce the explicit expression for the ghost number $k$ states that span the free $R$–modules $V_k^N$:
\begin{equation}
|W;\,a_1,\dots,a_k\rangle\;=\;W(Z)\,\eta_{-(N+1+a_1)}\cdots \eta_{-(N+1+a_k)}\,|0\rangle\,,
\qquad W\in R\,,
\end{equation}
with $\eta_X$'s anti-commuting (Grassmann) variables. The natural inner product is block–diagonal in ghost number and defined as follows
\bea
\label{eq:inner-product}
\langle W';\,a_1',\dots,a_k'|W;\,a_1,\dots,a_k\rangle
&=&\underbrace{\langle W'^\dagger\,W\rangle}_{\text{Wick contractions in }Z,Z^\dagger}
\;\times\;
\underbrace{\langle a_1',\dots,a_k' \mid a_1,\dots,a_k\rangle_{\eta}}_{\text{ghost Kronecker w/ signs}}\cr
&&
\eea
where the $\eta$-sector is orthonormal up to the canonical antisymmetric sign, and the $Z$-sector is computed with the rules for evaluating correlators in the free ring $R$. Equivalently, the $Z$-sector is computed using an analytic continuation in $N$. By this we mean that, whether one is working in the free ring or after analytically continuing to non-integer $N$, the trace relations are not imposed; instead, one simply applies Wick’s theorem using the rule (\ref{WickRule}).

Let $O(Z)$ be any ghost-number–zero gauge-invariant operator. To be concrete, it could be a short single trace operator, or a short single trace times a physical giant graviton represented as a Schur polynomial with $<N$ rows. Acting by left multiplication on $R$, $O(Z)$ preserves ghost number
\bea
O(Z): V_k^N \longrightarrow V_k^N,\qquad
O(Z)\big(W(Z)\otimes \omega\big)=(O(Z)W(Z))\otimes \omega\,.
\eea

To compute the transition amplitude involving an over maximal giant graviton, which corresponds to a trace relation, we must compute a matrix element using a $Q_b$–exact bra. Choose $|\Psi\rangle\in V_k^N$ and $|\Lambda\rangle\in V_{k+1}^N$ we consider the matrix element
\bea
\langle Q_b\Lambda \mid O(Z) \mid \Psi\rangle\;=\;
\sum_{\text{(matching ghost labels)}} \big\langle \big(\text{one }\chi_{(1^{N+1+\alpha})}(Z)\big)^\dagger \, \big(O(Z)\,W(Z)\big) \big\rangle\,.\nn\\
\eea
The right hand side is a sum of ordinary half-BPS two-point correlators in the free ring $R$, one for each surviving ghost match.

An interesting transition amplitude is the minimal non-trivial example which has $k=1$. In this case we can take
\bea
|\Psi\rangle =\eta_{-(N+1+b)}|0\rangle\,,\qquad |\Lambda\rangle=\eta_{-(N+1+a)}\!\wedge\!\eta_{-(N+1+b)}|0\rangle\,,
\eea
with $a\neq b$. Using the formulas above
\bea
\langle Q_b\Lambda \mid O(Z) \mid \Psi\rangle &=&\big\langle \chi_{(1^{N+1+a})}(Z^\dagger) \, O(Z) \big\rangle\,.
\eea
Choosing $O=\chi_{N-b}(Z)p_{a+1+b}(Z)$ yields the amplitude
\bea
\langle \chi_{(1^{N+1+a})}(Z)^\dagger \chi_{(1^{N-b})}(Z)p_{a+1+b}(Z)\rangle\,,
\eea
which is exactly what we compute by analytic continuation. Thus, we have a direct relation between analytic continuation in $N$ and the  ghost brane states of~\cite{Gaiotto:2021xce,Lee:2023iil,Lee:2024hef}.

We can now return to the results of Section \ref{sec:schurholog} and interpret them in the light of the above discussion. Recall formula (\ref{CPtracerel}), which reads
\be
\langle \det(\lambda\mathbf{1}-Z) \Tr(\bar{Z}^J)\rangle=-\lambda^{N-J}\frac{\Gamma(N+1)}{\Gamma(N-J+1)}\,.
\ee
For $N$ a positive integer, the above correlator vanishes for $J>N$. For non-integer $N$, the above correlator is non-zero and it corresponds to the transition between a ghost brane state located at a position determined by $\lambda$ in the bulk and a single trace chiral-primary operator.

Next, recall the formula (\ref{DetR}) describing the correlator between a sub-determinant operator and a general operator, which reads
\bea
\langle \chi_{(1^k)}(\lambda\mathbf{1}-Z)\chi_{(1^n)}(\bar{Z})\rangle&=&\frac{(-1)^n\lambda^{k-n}\,\Gamma(N+1)}{\Gamma(k-n+1)\Gamma(N-k+1)}\,.
\eea
To consider a trace relation we can take the integer $n$ to be greater than $N$. Since we have integer $k$, the Gamma function in the denominator sets this correlator to zero. However, for a maximal giant graviton corresponding to the determinant with $k=N$, we have
\bea
\langle \det(\lambda\mathbf{1}-Z) \chi_{(1^n)}(\bar{Z})\rangle&=&(-1)^n\lambda^{N-n}\frac{\Gamma(N+1)}{\Gamma(N-n+1)}\,.
\eea
After continuing $N$ to $\mathbb{C}$, $n$ can exceed $N$ and the correlator is non-zero. It corresponds to a transition between a maximal giant graviton with location in the bulk set by $\lambda$ and a ghost brane state.

Finally, recall the correlator (\ref{finalcorrelator}), which is given by
\bea
  \left< G\left|{{\rm Tr} (Z^J)\over \sqrt{JN^J}}\right|G \right> = \sqrt{\frac{\Gamma(N-k+J)}{JN^J\Gamma(N-k)}}\;.\label{finalcorrelator}
\eea
This expression gives a formula for one-point functions of chiral primaries, in a background representing trace relations where $k>N$, for non-integer N.
\section{Conclusions}\label{sec:Conclusions}
We have studied the analytic continuation of conformal field theory with gauge group $U(N)$ to arbitrary complex rank $N\in \mathbb{C}$. For positive integer $N$, trace relations appear, manifest in the fact that Schur polynomials with more than $N$ rows vanish. By contrast, when $N$ is not a positive integer, such Schur polynomials no longer vanish but instead define {\it evanescent states}. Developing this connection, we have proposed that ghost-brane states---bulk gravitational states dual to trace relations~\cite{Lee:2023iil}---should be identified with these evanescent states.

Building on this identification, we have shown that the inclusion of evanescent states leads to new expansions of the ${1\over 2}$-BPS index. These expansions are mathematically in complete agreement with the giant graviton expansion. The proposed connection has allowed us to compute transition amplitudes involving ghost-brane states and physical giant graviton states, which we demonstrated to coincide with analytically continued CFT amplitudes. 

Taken together, our results establish a concrete bridge between analytic continuation in $N$ and bulk constructions involving giant gravitons, ghost branes, and thimbles~\cite{Gaiotto:2021xce,Lee:2023iil,Lee:2024hef}. They highlight analytic continuation as a powerful tool for probing finite-$N$ constraints and for uncovering novel aspects of the gauge/gravity correspondence. 

Several directions for future work remain open. A natural extension is to move beyond the half-BPS sector to less supersymmetric or even non-supersymmetric sectors, where the structure of trace relations becomes more intricate. In the multi-matrix setting the Schur polynomial basis is replaced by the restricted Schur polynomials (see~\cite{deMelloKoch:2024sdf} for a pedagogical overview). Here the trace relations appear as restricted Schur polynomials labelled by Young diagrams with more than $N$ rows, which again become evanescent states once we continue in $N$. It should be straightforward to adapt our analysis to this setting and derive giant-graviton-like expansions for the free theory. On the bulk side, one would correspondingly expect ghost-brane states associated to each of these evanescent states. Relevant works for this direction include \cite{Eleftheriou:2023jxr,Eleftheriou:2025lac}.

Another natural avenue is to develop the $SO(N^*)$ and $Sp(N^*)$ generalizations and to clarify how the $N \to -N$ relation between these gauge groups~\cite{Ramgoolam:1993hh,Caputa:2013vla} manifests after analytic continuation. For these theories, the analogues of Schur polynomials were constructed in~\cite{Caputa:2013hr,Caputa:2013vla}, while the analogues of restricted Schur polynomials were developed in~\cite{Kemp:2014apa,Kemp:2014eca}. This suggests that our analysis can be straightforwardly extended to obtain giant-graviton-like expansions in both the single- and multi-matrix settings. Related results, obtained using different methods, appear in~\cite{Eleftheriou:2025lac}.

We have studied evanescent operators constructed using scalar fields. Restricted Schur polynomials constructed for composite operators that include fermions~\cite{deMelloKoch:2012sie,Berenstein:2019esh} and gauge fields~\cite{deMelloKoch:2011vn} have been studied. These provide a natural starting point to study more general classes of evanescent operators constructed using fermions and gauge fields.

It would be interesting to investigate further the connection between evanescent states and the thimble formalism~\cite{Lee:2024hef}, with the goal of verifying their precise bulk duals and clarifying the geometric meaning of the giant graviton expansion.

The central objects in our discussion are the trace relations, which have played an important role in a number of recent works. In the study of ${1\over 16}$-BPS states in ${\cal N}=4$ super Yang--Mills theory, new {\it fortuitous} states have been discovered~\cite{Chang:2022mjp,Choi:2022caq,Chang:2023zqk,Choi:2023znd,Choi:2023vdm}. The BPS saturation of these states relies entirely on the presence of trace relations. It is therefore natural to ask whether analytic continuation in $N$ can provide additional insight into the construction and properties of such fortuitous states. Furthermore, the papers \cite{deMelloKoch:2025ngs,deMelloKoch:2025rkw,deMelloKoch:2025cec,deMelloKoch:2025qeq} have explicitly solved the trace relations to obtain a set of generators that freely generate the complete ring of gauge-invariant operators. This analysis has yielded important lessons about the UV structure of the gravity Fock space. Once again, it would be interesting to explore what further insights might be gained by considering analytic continuation in $N$. 

Finally, and perhaps more speculatively, the ideas developed in this paper may find useful application in the context of the dS/CFT proposals such as~\cite{Hikida:2021ese}, where one is also faced with non-trivial continuation of the CFT parameters, and their non-trivial gravity interpretation. Indeed, the large central charge limit $k \to -N$ of the $SU(N)$ WZW model corresponds to the classical limit of $SU(N)$ higher-spin gravity on $S^3$ in this setup. More generally, understanding the connection with developments around \cite{Witten:2010cx} may also be an interesting avenue and we leave these intriguing questions for future investigation.

\bigskip
\noindent {\bf \large Acknowledgments}
\\
We are grateful to Sameer Murthy, Piotr Sułkowski, Bo Sundborg, Balt van Rees, Jaco Van Zyl and Tadashi Takayanagi for discussions, and to Ji Hoon Lee, Shinji Hirano and Adolfo Holguin for discussions and comments on the draft. This work is supported by the ERC Consolidator grant (number: 101125449/acronym: QComplexity).  Views and opinions expressed are however those of the authors only and do not necessarily reflect those of the European Union or the European Research Council. Neither the European Union nor the granting authority can be held responsible for them. PC is supported by the NCN Sonata Bis 9 2019/34/E/ST2/00123 grant.
The work of RdMK is supported by a start up research fund of Huzhou University, a Zhejiang Province talent award and by a Changjiang Scholar award.
\appendix

\bibliographystyle{nb}
\bibliography{mainv2}
\end{document}